\documentclass[aps,prl,twocolumn,showpacs,superscriptaddress,groupedaddress,reprint]{revtex4-1}
\usepackage[hidelinks]{hyperref} 
\usepackage{comment}
\usepackage{caption} 
\usepackage{dsfont}
\usepackage{amsthm}
\usepackage{amsmath}
\usepackage{amssymb}
\usepackage{esint}
\usepackage{graphicx, accents}
\usepackage{mathrsfs}             
\usepackage{amssymb}
\usepackage{epstopdf}
\usepackage{textpos}
\usepackage{lipsum}
\usepackage{color}
\usepackage{blindtext}
\makeatletter
\DeclareRobustCommand{\cev}[1]{%
  \mathpalette\do@cev{#1}%
}
\newcommand{\do@cev}[2]{%
  \fix@cev{#1}{+}%
  \reflectbox{$\m@th#1\vec{\reflectbox{$\fix@cev{#1}{-}\m@th#1#2\fix@cev{#1}{+}$}}$}%
  \fix@cev{#1}{-}%
}
\newcommand{\fix@cev}[2]{%
  \ifx#1\displaystyle
    \mkern#23mu
  \else
    \ifx#1\textstyle
      \mkern#23mu
    \else
      \ifx#1\scriptstyle
        \mkern#22mu
      \else
        \mkern#22mu
      \fi
    \fi
  \fi
}

\makeatother
\DeclareMathOperator{\tr}{tr}

\newcommand{\C}{\mathbb{C}}

\newcommand{\ket}[1]{{\vert{#1}\rangle}}
\newcommand{\bra}[1]{{\langle{#1}\vert}}
\newcommand{\state}[1]{{\ket{#1}\bra{#1}}}
\newcommand{\abs}[1]{\vert{#1}\vert}
\newcommand{\ceqref}[1]{Eqn.\! (\ref{#1})}
\newcommand{\wh}{\widehat}
\newcommand{\ot}{\otimes}

\begin{document}
\begin{textblock*}{24cm}(10cm,-1cm)
\fbox{\footnotesize $\text{SU-ITP-17/03}$}
\end{textblock*} 
\title{Out-of-time-order Operators and the Butterfly Effect}
\date{\today}
\author{Jordan S. Cotler}
\affiliation{Stanford Institute for Theoretical Physics, Stanford University, Stanford, CA 94305}
\author{Dawei Ding}
\affiliation{Stanford Institute for Theoretical Physics, Stanford University, Stanford, CA 94305}
\author{Geoffrey R. Penington}
\affiliation{Stanford Institute for Theoretical Physics, Stanford University, Stanford, CA 94305}

\begin{abstract}
Out-of-time-order (OTO) operators have recently become popular diagnostics of quantum chaos in many-body systems. The usual way they are introduced is via a quantization of classical Lyapunov growth, which measures the divergence of classical trajectories in phase space due to the butterfly effect.  However, it is not obvious how exactly they capture the sensitivity of a quantum system to its initial conditions beyond the classical limit.  In this paper, we analyze sensitivity to initial conditions in the quantum regime by recasting OTO operators for many-body systems using various formulations of quantum mechanics.  Notably, we utilize the Wigner phase space formulation to derive an $\hbar$--expansion of the OTO operator for spatial degrees of freedom, and a large spin $1/s$--expansion for spin degrees of freedom.  We find in each case that the leading term is the Lyapunov growth for the classical limit of the system and argue that quantum corrections become dominant at around the scrambling time, which is also when we expect the OTO operator to saturate.  We also express the OTO operator in terms of propagators and see from a different point of view how it is a quantum generalization of the divergence of classical trajectories.
\end{abstract}


\maketitle


\section*{Introduction}
Quantum chaos attempts to generalize well-established classical diagnostics of chaos to quantum-mechanical systems~\cite{devaney1992first,hasselblatt2003first,zurek1994decoherence,cucchietti2003decoherence}. Although classical diagnostics are well-understood, finding satisfying quantum diagnostics is an ongoing field of research~\cite{gutzwiller2013chaos,haake2013quantum,stockmann2007quantum,wimberger2014nonlinear}.  One reason is that na\"ive quantum generalizations of classical measures of chaos often prove unsatisfactory.

For instance, we can diagnose classical chaos by looking at the sensitivity of dynamics to initial conditions. That is, for chaotic systems, two nearby initial states will diverge quickly under time evolution. This phenomenon is commonly known as the butterfly effect. We could try to generalize this notion to the quantum case in the following way. Given a chaotic quantum system, for an initial state $|\Psi\rangle$ and some perturbed state $|\Psi'\rangle$, one might na\"ively expect that their inner product diminishes quickly with time. However, due to the unitarity of time evolution, the inner product actually stays \textit{constant}:
$$\langle \Psi'|U^\dagger(t)\, U(t)|\Psi\rangle = \langle \Psi'|\Psi\rangle\,.$$
Hence we cannot characterize quantum chaos by looking at the evolution of the overlap between states \footnote{Other notions of distance between states in the Hilbert space such as the relative complexity~\cite{brown2017second} have the potential to characterize chaos, but are hard to work with directly.}. 

There has been greater success in characterizing quantum chaos by considering the time evolution of observables instead of states. To do this, we first mathematically express the butterfly effect as the sensitivity of a system's trajectory $x(t)$ to its initial condition $x_0$. Specifically, a system exhibits the butterfly effect if
\begin{equation}
  \left|\frac{\partial x(t)}{\partial x_0}\right| \sim e^{\lambda t}\,,
  \label{}
\end{equation}
where $\lambda > 0$ is known as a \textit{Lyapunov exponent}. To generalize this notion to quantum systems, we re-write the sensitivity as a Poisson bracket:
\begin{equation}
\label{classical1}
\frac{\partial x(t)}{\partial x_0} = \{x(t), \, p_0\}_{\text{Poisson}}\,,
\end{equation} 
where $p_0$ is the initial momentum. We can now proceed by canonical quantization to obtain the quantity 
$$\frac{1}{i\hbar}[\widehat{x}(t), \widehat{p}]\,.$$
We are often interested in expectation values of this observable with respect to some density matrix $\rho$, but cancellations could occur due to terms with differing sign.  Hence, it is natural to take the norm squared of this operator:
\begin{equation}
\label{OTO1}
\frac{1}{i \hbar}[\widehat{x}(t), \widehat{p}]\cdot -\frac{1}{i \hbar}[\widehat{x}(t), \widehat{p}]^\dagger = -\frac{1}{\hbar^2}[\widehat{x}(t), \widehat{p}]^2.
\end{equation}
The above is an example of an out-of-time-order (OTO) operator, which first appeared in \cite{larkin1969quasiclassical} and was studied in~\cite{almheiri2013black,shenker2014black,shenker2014multiple,shenker2015stringy,kitaev2014hidden,maldacena2016bound,roberts2015localized} in the context of black hole physics and large-$N$ quantum field theories.  It is so named since it contains terms that are not time-ordered such as $\widehat{x}(t) \, \widehat{p} \, \widehat{x}(t)\, \widehat{p}$.  Note that in recent literature, any operator of the form $[\widehat{W}(t), \widehat{V}]\cdot [\widehat{W}(t), \widehat{V}]^\dagger$ is referred to as an OTO operator (see, for instance,~\cite{swingle2016measuring,halpern2017jarzynski}).  The corresponding many-body, higher dimensional version of \ceqref{OTO1} is given by
\begin{equation}
  -\frac{1}{\hbar^2} \left[ \wh x_i(t), \wh p_j \right]^2,
  \label{ManyBodyGeneralization1}
\end{equation}
where $x_i, p_j$ are the position and momentum of the $i$th and $j$th coordinate, respectively. Eqn. \!(\ref{ManyBodyGeneralization1}) quantifies the sensitivity of the position of the $i$th coordinate to the initial position of the $j$th coordinate.

It is not obvious how exactly the OTO operator characterizes the sensitivity of a quantum system to its initial conditions beyond the classical limit. Indeed, the OTO is known to saturate at the scrambling or Ehrenfest time, while $\left|\frac{\partial x(t)}{\partial x_0}\right|$ of a chaotic system should grow exponentially forever. In this paper, we consider the OTO operator in different formulations of quantum mechanics to make explicit how it quantifies sensitivity to initial conditions -- the butterfly effect.  We find that using the Wigner phase space representation, an OTO operator can be written as a semiclassical expansion with the leading classical term equal to $\left( \frac{\partial x(t)}{\partial x_0} \right)^2$.  We further argue that the quantum corrections become dominant around the scrambling time, providing a heuristic explanation for the ending of the exponential growth at that time scale.  As an example, we consider the BFSS matrix model and correctly predict its scrambling time.  Additionally, we rewrite OTO operators in terms of propagators in the Schr\"odinger formulation and see in another way how they capture sensitivity to initial conditions.  Our analysis is repeated for an OTO operator for spin degrees of freedom. 

\section*{Spatial Degrees of Freedom}

\subsection*{Phase Space Representation}


In this section we use the Wigner phase space formulation~\cite{wigner1932quantum,groenewold1946principles,moyal1949quantum} to study OTO operators for spatial degrees of freedom, that is, those of the form given in \ceqref{ManyBodyGeneralization1}. This formulation is particularly useful for studying the quantum-classical correspondence and we will use it to obtain a semiclassical expansion of the OTO. For a nice overview of this formulation of quantum mechanics, see~\cite{curtright2013concise}.

Recall that in classical mechanics on phase space
we have a time-dependent probability distribution $\rho^{\text{class}}(\vec{x},\vec{p},t)$ over the phase space variables which satisfies Hamilton's equations of motion.  Phase space can be generalized to the quantum mechanical setting via the Weyl correspondence~\cite{weyl1927quantenmechanik}: given a density matrix $\widehat{\rho}(t)$ of $N$ particles in $d$ spatial dimensions, we can compute the associated Wigner distribution on a $2\,d\,N$--dimensional phase space with coordinates $(\vec{x},\vec{p})$:
\begin{align}\label{WignerTrans1}
\rho(\vec{x},\vec{p},t) &\equiv \mathcal{W}_{\widehat{\rho}(t)}(\vec{x},\vec{p})  \nonumber \\
&\equiv \frac{1}{(\pi \hbar)^{dN}}\int_{\mathbb{R}^{d N}} d\vec{y} \, \langle \vec{x} + \vec{y}|\,\widehat{\rho}(t)\,|\vec{x}-\vec{y}\rangle \, e^{-2i \vec{p}\cdot \vec{y}/\hbar}\,.
\end{align}
The Wigner function is a quasiprobability distribution, meaning that it integrates to unity but can take on negative values.  In fact it is only negative on $\hbar$-scale cells of phase space, and is otherwise positive -- it is much like a classical phase space distribution with small quantum effects causing it to be negative in some places~\cite{curtright2013concise}.  The Wigner distribution satisfies the correspondence principle~\cite{wang1986classical}
\begin{equation}
\lim_{\hbar \to 0} \rho(\vec{x},\vec{p},t) = \rho^{\text{class}}(\vec{x},\vec{p},t),
\end{equation}
where intuitively the quasiprobability distribution tends to a classical probability distribution evolving under classical Hamiltonian dynamics as the negative regions approach zero size. Note that what is physically meant by taking $\hbar \to 0$ is that $\hbar$ is small compared to the length and momentum scales considered.  

The map given in \ceqref{WignerTrans1} for density matrices can be applied to general operators to obtain the corresponding phase space functions in the Wigner representation. It induces a non-commutative product known as the \textit{Moyal star product} given by
\begin{align}
\label{StarProduct1}
& A(\vec{x},\vec{p}) \star B(\vec{x},\vec{p}) \nonumber \\
& \qquad \equiv A(\vec{x},\vec{p}) \, \exp\left[\frac{i\hbar}{2}\left(\cev{\partial}_{\vec{x}}\cdot \vec{\partial}_{\vec{p}} -  \cev{\partial}_{\vec{p}} \cdot \vec{\partial}_{\vec{x}} \right) \right] \, B(\vec{x},\vec{p}).
\end{align}
The Moyal star product is induced by the Wigner representation in the sense that $\mathcal{W}_{\widehat{A}} \star \mathcal{W}_{\widehat{B}} = \mathcal{W}_{\widehat{A}\, \widehat{B}}$.  From \ceqref{StarProduct1}, we see that the Moyal star product naturally gives an $\hbar$--expansion. It is also convenient to define the phase space representation of the commutator called the \textit{Moyal bracket}:
\begin{align}
\label{MoyalBracket}
\{\!\{A, B\}\!\} &\equiv \frac{1}{i \hbar}(A \star B - B \star A) \\
&= \{A,B\}_{\text{Poisson}} \,\,+\mathcal{O}(\hbar^2).
\end{align}
In the limit $\hbar \to 0$, the Moyal bracket becomes the Poisson bracket, thus manifestly satisfying the correspondence principle.

We will now rewrite OTO operators as phase space functions in the Wigner formalism:
\begin{equation}
  -\frac{1}{\hbar^2} \left[ \wh x_i(t), \wh p_j \right]^2 \mapsto \{\!\{X_i(\vec{x},\vec{p},t), P_j(\vec{x},\vec{p},0)\}\!\}^{\star 2}\,.
  \label{OTOtoWigner1}
\end{equation}
Note that although the phase space functions for position and momentum have the initial conditions $X_i(\vec{x},\vec{p},t=0) = x_i$ and $P_j(\vec{x},\vec{p},t=0) = p_j$, they do \textit{not} evolve by Hamilton's classical equations of motion: instead, they are given by the solutions to the classical equations of motion with $\hbar$ corrections~\cite{curtright2013concise}. For instance,
\begin{equation}
\label{MoyalExpansion1}
  X_i(t) = x^{\mathrm{cl}}_i (t) + \sum_{k=1}^{\infty} \hbar^{2k} x_i^{(k)}(t)
\end{equation}
is the solution to the Moyal equation of motion
\begin{equation}
\dot X_i (t) = \{\!\{H, X_i (t)\}\!\} = \{H, X_i (t)\} + \mathcal{O}(\hbar^2)\,.
\end{equation}
These $\hbar$ corrections depend on the dynamics under consideration and are not easily characterized for general systems. As a result, we will leave the corrections packaged implicitly into $X_i(t)$, since we would like to derive an expression \textit{independent} of the specific dynamics of the system.  That is, we will provide an $\hbar$-expansion for the OTO but will suppress the $\hbar$ dependence of $X_i(t)$.

Using $d\,N$-dimensional multi-index notation, we obtain the expansion:
\begin{widetext}
\begin{equation}
\label{OTOexpansion1}
\{\!\{X_i(\vec{x},\vec{p},t), P_j(\vec{x},\vec{p},0)\}\!\}^{\star 2} = \sum_{\ell=0}^{\infty} \frac{1}{(2\ell)!} \left( \frac{i \hbar}{2} \right)^{2\ell} \sum_{\abs{\vec m} =2\ell}^{} \binom{2\ell}{\vec m} \sum_{\vec n \le \vec m} (-1)^{\abs{\vec n}}\binom{\vec m}{\vec n}  \left( \partial_x^{\vec{n}} \partial_p^{\vec{m} - \vec{n}} Y_{ij}(t) \right) \left( \partial_x^{\vec{m} - \vec{n}} \partial_p^{\vec{n}} Y_{ij}(t) \right)
\end{equation}
\end{widetext}
where $|\vec{v}| \equiv v_1 + \cdots + v_{d\,N}$ and $Y_{ij}(t)\equiv \frac{\partial X_i(t)}{\partial x_j}$\,. 

To better understand Eqn. \!(\ref{OTOexpansion1}), let us consider the simplified setting of a single particle in one dimension.  In this case, the leading terms of Eqn. \!(\ref{OTOexpansion1}) reduce to
\begin{equation} \label{OTOexpansion2}
\left(\frac{\partial X(t)}{\partial x}\right)^2 - \frac{\hbar^2}{4} \,\mathrm{det}\left[\mathrm{Hess}\left(\frac{\partial X(t)}{\partial x} \right)\right] + \mathcal{O}(\hbar^4)
\end{equation}
where $\text{Hess}(\,\cdot\,)$ is the Hessian on the two-dimensional phase space (i.e., the matrix of second derivatives with respect to the initial conditions $X(0) = x$ and $P(0) = p$). In Appendix \ref{sec:manyWignerOTO}, we detail how the $\mathcal{O}(\hbar^2)$ term in the many-body case can be expressed in terms of a higher-dimensional analog of the Hessian. As expected, the first term in Eqn.\,(\ref{OTOexpansion2}) contains the square of \ceqref{classical1} since $X(t) = x^{\text{cl}}(t) + \mathcal{O}(\hbar^2)$, and so we manifestly reproduce classical chaotic behavior as $\hbar \to 0$.  The $\mathcal{O}(\hbar^2)$ corrections in Eqn.\! (\ref{OTOexpansion1}) are interesting because they correspond to \textit{higher} derivatives with respect to the initial conditions of $X(t)$. In fact, $\mathcal{O}(\hbar^{2k})$ terms contain products of $(2k+1)$th derivatives with respect to initial conditions. 

\subsection*{Saturation of Chaos}

Here we argue that the semiclassical expansion of the OTO operator can explain the operator's early time growth and heuristically justify the timescale at which this growth saturates -- namely the scrambling time. (In the original literature on semiclassical expansions, this is known as the Ehrenfest time, but we shall use scrambling time throughout this paper.)  We will focus mostly on the expansion for a single particle in one dimension for simplicity, but the generalization to more dimensions and particles is straightforward. Note that although by definition a single classical particle in one dimension is always integrable, it can still be exponentially sensitive to initial conditions and thus is still instructive.

The leading term $\left(\frac{\partial x^{\mathrm{cl}}(t)}{\partial x}\right)^2$ in the expansion of the OTO operator in the Wigner representation simply describes the sensitivity of the position at time $t$ to the initial condition in the classical limit. For a chaotic system, it should grow exponentially for all time due to the butterfly effect:
\begin{align}
\left(\frac{\partial x^{\mathrm{cl}}(t)}{\partial x}\right)^2 \sim e^{2 \lambda t}.
\end{align}
However, expectation values of the OTO operator are known to initially grow exponentially but then saturate to a finite value~\cite{hashimoto2017out}. This suggests that the OTO operator stops growing and saturates because the expansion is no longer dominated by the leading classical term.

We now make a heuristic argument that the classical term stops dominating at the time known as the scrambling time, when the exponential growth ends. The argument is similar to the reasoning used in~\cite{zurek1994decoherence} to explain the evolution of the Wigner representation of the density matrix becoming dominated by non-classical terms after exactly the same timescale, but we consider the evolution of operators rather than states. The two arguments are related by the usual duality in quantum mechanics between the Schr\"odinger and Heisenberg pictures of time evolution. Both points of view describe the same physical behavior and so both must contain the same phenomena.

The two leading quantum corrections are of order $\hbar^2$ and are
\begin{align} \label{eq:corrections}
2 \hbar^2 \frac{\partial x^{(1)}(t)}{\partial x} \frac{\partial x^{\mathrm{cl}} (t)}{\partial x}- \frac{\hbar^2}{4} \,\mathrm{det}\left[\mathrm{Hess}\left(\frac{\partial x^{\mathrm{cl}}(t)}{\partial x} \right)\right]
\end{align}
where $\hbar^2 x^{(1)}(t)$ is the leading quantum correction to the Moyal trajectory $X(t)$ as per Eqn. (\ref{MoyalExpansion1}).

We first consider the correction in the right half of Eqn.\! (\ref{eq:corrections}), which comes from higher derivatives of the classical trajectory.  We shall refer to this as the ``Hessian correction." Consider the quantity
$$\mathcal{A}(f) = \sqrt{\frac{f^2}{\mathrm{det}\, \text{Hess}(f)}}= \sqrt{\frac{f}{\frac{\partial^2 f}{\partial z_1^2}} \frac{f}{\frac{\partial^2 f}{\partial z_2^2}}}$$
where $z_1, z_2$ are local coordinates on phase space that diagonalize the Hessian. Up to order one factors, $\mathcal{A}(f)$ characterizes the maximum possible area of an ellipse (with $z_1,z_2$ forming the principal or minor axes) within which the second-order terms in the Taylor series of $f$ with respect to $z_1, z_2$ are small compared to $f$ itself. The Hessian correction is small compared to the leading classical term when
\begin{align}
\mathcal{A}\left(\frac{\partial x^{\mathrm{cl}}(t)}{\partial x}\right) \gg \hbar \, .
\end{align}
Hence, the Hessian correction can be ignored when there is an ellipse larger than the Planckian area around the initial conditions $(x,p)$ where $\frac{\partial x^{\mathrm{cl}}(t)}{\partial x}$ is well-approximated by a Taylor expansion truncated to first order.

Liouville's theorem states that classical time evolution on phase space is area-preserving.  However for chaotic systems, an ellipse will get mapped to increasingly warped and complicated regions of phase space at long times.  This mapping should become highly non-linear for all ellipses of area $A$ when
\begin{align}
e^{\lambda t} \gg \frac{L_x L_p}{A}
\end{align}
where $L_x$ and $L_p$ are the characteristic length and momentum scales at the energy determined by the initial conditions $(x,p)$. $L_x L_p$ is therefore the effective accessible area of phase space.  This suggests that $\mathcal{A}\left(\frac{\partial x^{\mathrm{cl}}(t)}{\partial x}\right) \lesssim \hbar$ and the expansion of the OTO operator is no longer dominated by its leading term when
\begin{align} e^{\lambda t} \gtrsim \frac{L_x L_p}{\hbar}\,.
\end{align}
This timescale is exactly the scrambling time.

We now consider the correction in the left half of \ceqref{eq:corrections}, which we shall refer to as the ``trajectory correction." It was argued in ~\cite{zurek1994decoherence} when considering the evolution of states in the Wigner picture that smooth functions $f$ on phase space should evolve approximately classically for chaotic systems, that is, $f(t) \sim f^\text{cl}(t)$, until the scrambling time
\begin{align}
t_{\mathrm{scr}} \sim \lambda^{-1} \log\left(\frac{L_x L_p}{\hbar}\right)
\end{align}
where quantum corrections begin to dominate. We shall briefly review this argument for the specific case of $X(t)$. The leading quantum correction to the Moyal equation of motion occurs at order $\hbar^2$ and contains two extra derivatives with respect to each of $x$ and $p$.  By dimensional analysis, we might na\"{i}vely assume that this means
\begin{align}
\frac{\hbar^2\,\dot x^{(1)} (t)}{\dot x^{\mathrm{cl}} (t)} \sim \frac{\hbar^2}{L_x^2 L_p^2}\,.
\end{align}
However, in each term of the leading correction to Moyal's equation, there are two derivatives acting on $X(t)$ and two others acting on the Hamiltonian.  The two derivatives acting on $X(t)$ expose its exponential sensitivity to initial conditions by producing two factors of $e^{\lambda t}$ (at least as long as it is evolving approximately classically). As a result we instead expect
\begin{align}
 \frac{\hbar^2 \,\dot x^{(1)} (t)}{\dot x^{\mathrm{cl}} (t)} \sim \frac{\hbar^2}{L_x^2 L_p^2} \, e^{2 \lambda t}\,,
\end{align}
and so, just as for the Hessian correction, we should expect that the trajectory correction will become significant at the scrambling time. Furthermore, although we only considered the leading corrections at order $\hbar^2$, a similar analysis holds for all other orders. As a result, the semiclassical expansion will entirely break down at the scrambling time.  

The classical term in the OTO operator expansion is given at the scrambling time by
\begin{align} \left(\frac{\partial x_{\mathrm{cl}}(t_{\text{scr}})}{\partial x}\right)^2 \sim e^{2 \lambda t_{\text{scr}}} \sim \frac{(L_x L_p)^2}{\hbar^2}\,.
\end{align}
Note that we cannot directly predict the saturation of the OTO at the scrambling using the Wigner expansion. Instead, we can only see the end of the semiclassical exponential growth of the OTO. Going beyond the scrambling time and seeing the saturation will require different techniques and approximations.

It is interesting that both corrections naturally become large at the same timescale. Essentially, the behavior is classical until the scrambling time when the phase space description becomes dominantly quantum. Not only does the semiclassical expansion of the OTO break down, but even the Moyal trajectory $X(t)$ ceases to have any direct physical meaning.

With multiple dimensions or particles, we have to sum over Hessian corrections to the OTO operator for all pairs of conjugate dimensions, as discussed in Appendix \ref{sec:manyWignerOTO}. Performing a linear canonical transformation so that the Lyapunov growth is diagonal, we see that the quantum correction corresponding to the pair with the largest Lyapunov exponent $\lambda_{\max}$ will be the first to become significant. Similarly, the largest corrections to the Moyal trajectory will come from derivatives associated with the same conjugate dimensions (call them $x,p$). The scrambling time is therefore
\begin{align}
\label{scrambMax}
t_{\mathrm{scr}} = \lambda_{\mathrm{max}}^{-1} \log \frac{L_x L_p}{\hbar}\,.
\end{align}
This is consistent with results in the literature~\cite{maldacena2016bound}. We note that although for one particle in one dimension $L_x L_p$ was the accessible area of phase space, this is no longer true for higher dimensions. This is because the only part of phase space that determines the scrambling time is the two-dimensional subspace involving the largest Lyapunov exponent. All the additional dimensions are effectively irrelevant for the breakdown of semiclassical behavior.

There is recent interest in studying chaos for systems with a large number of matrix degrees of freedom.  One such system is the BFSS matrix model~\cite{banks1997m}, which is a chaotic quantum system of great importance in quantum gravity.  Chaos in the classical BFSS model was studied in~\cite{gur2015chaos}. It was shown that it is necessary to take the 't Hooft limit  for the largest Lyapunov exponent to converge to a fixed value at large $N$, the size of the matrices. 

In this limit, the ratio of the product of the accessible length and momentum scales to the Planck scale grows linearly with $N$. This is easiest to see if you take the 't Hooft limit by making the Planck scale $\frac{\hbar}{N}$ rather than $\hbar$, while keeping the coupling fixed. If you take the more traditional, but equivalent, approach of absorbing the factor of $\frac{1}{N}$ into the coupling, the Planck scale will be fixed but you instead get growth in the accessible size of phase space. Since the ratio $\frac{L_x L_p}{\hbar}$ is a dimensionless, physical quantity, it is the same from both points of view. If we apply our arguments to this specific example, we find correctly that the scrambling time $t_{\text{scr}} \sim \lambda_{\mathrm{max}}^{-1}\log (N/\hbar)$.

It is somewhat surprising at first glance why the scrambling time for the BFSS model and other fast-scrambling, large $N$ quantum systems should be proportional to $\log \left(N/\hbar\right)$ when the entropy grows as $N^2 \log\left(1/\hbar \right)$. Why should the scaling of the scrambling time be so different from that of the entropy when we increase the number of quantum states by adding more matrix degrees of freedom?
Our approach makes the intuition behind this much clearer: even though the full volume of phase space grows exponentially with $N$, the area of the two-dimensional subspace involving the largest Lyapunov exponent only grows linearly with $N$ in units of the Planck scale in the 't Hooft limit. As we discussed above, it is the area of this subspace that determines the scrambling time, rather than the entire volume of phase space. 

\subsection*{Propagator Representation}
Now we will further study the structure of OTO operators by expressing them in terms of propagators in the standard Schr\"{o}dinger formalism.  We will find that a natural kernel-like quantity arises which we will refer to as the \textit{chaos kernel}. The OTO operator can be interpreted roughly as the sensitivity of the first moment of the chaos kernel to initial conditions.

We first fix some notation for the propagator:
\begin{equation}
  K(\vec x, \vec y, t) \equiv \bra{\vec y} U \ket{\vec x},
  \label{}
\end{equation}
where $\ket{\vec x}, \ket{\vec y}$ are the position eigenstates of $N$ particles in $d$ spatial dimensions and $U$ is the time evolution operator with time parameter $t$. 
Then, we define the chaos kernel as
\begin{equation}
  C(\vec x,\vec y,\vec z,t) \equiv K(\vec x, \vec y, t) K^*(\vec z, \vec y,t)\,.
  \label{cKernel}
\end{equation}
This is the probability amplitude of $\vec x \to \vec y$, i.e., moving forward in time, multiplied by that of $\vec y \leftarrow \vec z$, i.e., moving backward in time. The chaos kernel can be imagined as the amplitude of a spacetime arc, as shown in Fig.\! \ref{fig:arc}.
\begin{center}
  \includegraphics[width=0.5\linewidth]{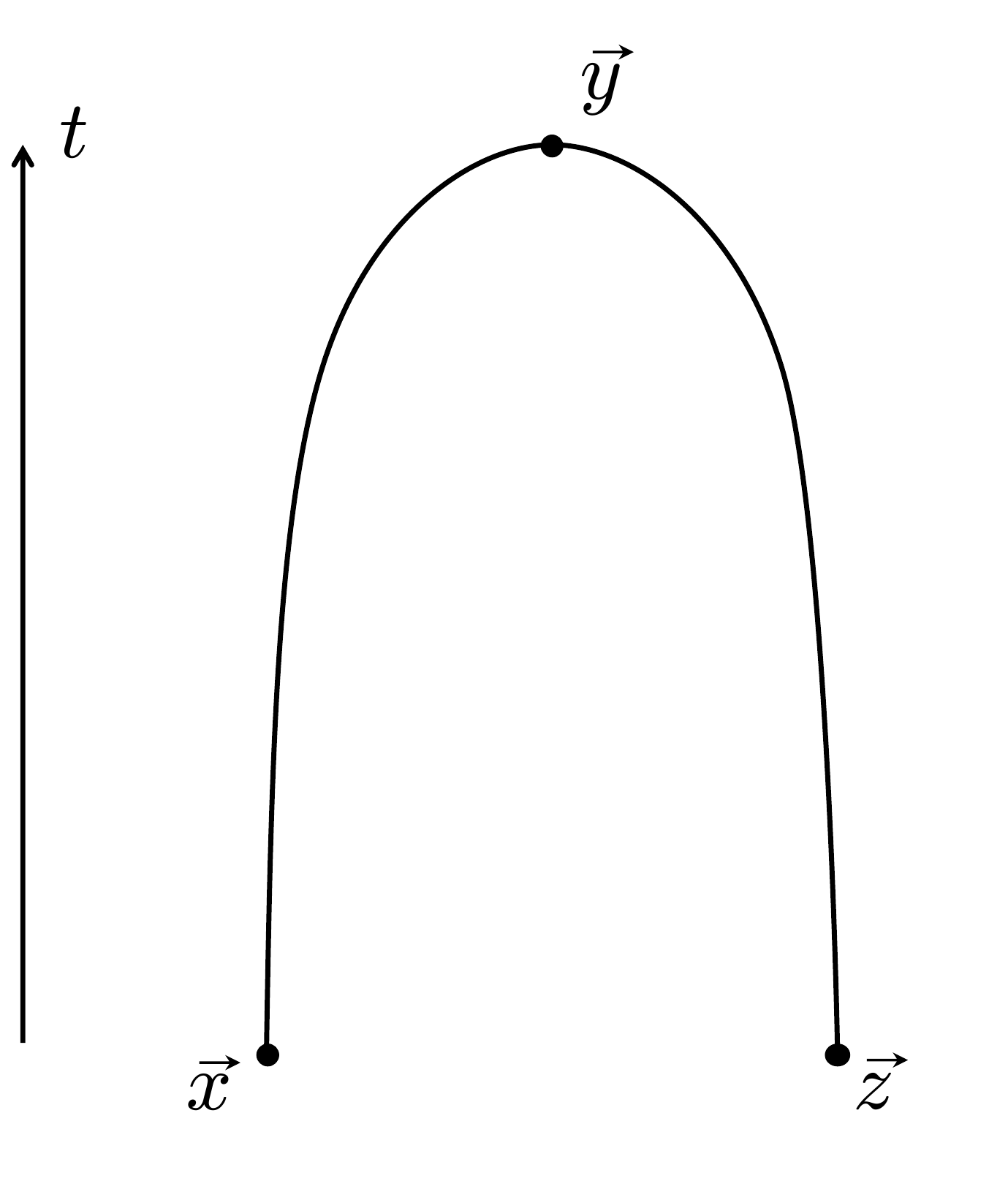}
  \captionof{figure}{An arc in (coordinate space)$\times$(time). $C(\vec x, \vec y, \vec z, t)$ is the corresponding probability amplitude.}
  \label{fig:arc}
\end{center}
Let $\mu_i(\vec x,\vec z,t)$ be the first moment of the chaos kernel with respect to the $i$th intermediate position coordinate:
\begin{equation}
  \mu_i(\vec x,\vec z,t) \equiv \int d\vec{y} \,\, y_i \, C(\vec x, \vec y, \vec z,t)\,.
  \label{1stMom}
\end{equation}
Physically, $\mu_i$ is the first moment of the $i$th intermediate position coordinate averaged over all spacetime arcs starting at $\vec x$ and ending at $\vec z$. We can then write the many-body OTO operator as
\begin{align}
   &\!\!\!\!-\frac{1}{\hbar^2}[\widehat{x}_i(t),\widehat{p}_j]^2 =  \int d\vec{x}_1 \, d\vec{x}_2 \, d\vec{x}_3 \,\,\ket{\vec x_1} \bra{\vec x_3} \qquad \nonumber \\
  & \left. \times \frac{\partial}{\partial x_{21,j}^+} \mu_i(\vec x_2, \vec x_1,t) \right \vert_{x_{21,j}^-} \left. \frac{\partial}{\partial x_{32,j}^+} \mu_i(\vec x_3, \vec x_2, t) \right \vert_{x_{32,j}^-},
  \label{OTOkernel}
\end{align}
where $x_{ab,j}^\pm \equiv \frac{1}{2} \left( x_{a,j} \pm x_{b,j} \right)$ and $x_{a,j}$ is the $j$th coordinate of $\vec x_a$. Each partial derivative can be physically interpreted as the sensitivity of $\mu_i$ to changes in the average position of the two ends of the arc while keeping the difference in the positions of the ends constant. We can imagine this change by a rigid 1D-motion of a rod connecting the two positions.  Note that the expression is manifestly Hermitian since
\begin{equation}
  \left. \frac{\partial}{\partial x_{21,j}^+} \mu_i(\vec x_2, \vec x_1,t) \right \vert_{x_{21,j}^-}  = \left(\frac{\partial}{\partial x_{2, j}} + \frac{\partial}{\partial x_{1, j}} \right) \mu_i(\vec x_2, \vec x_1,t),
  \label{}
\end{equation}
and $\mu_i^*(\vec x_2, \vec x_1 ,t ) = \mu_i(\vec x_1, \vec x_2 , t)$.

\ceqref{OTOkernel} captures the sensitivity of the quantum state at time $t$ to its initial conditions, just as \ceqref{classical1} does for a classical system. If we take $\vec x_1= \vec x_2$ for simplicity, the relationship between \ceqref{OTOkernel} and the butterfly effect becomes especially clear:
\begin{align}
  &\left. \left(\left. \frac{\partial}{\partial x_{21,j}^+} \mu_i(\vec x_2, \vec x_1, t) \right \vert_{x_{21,j}^-} \right)\right\vert_{\vec x_2 = \vec x_1 =\vec x}  \nonumber \\ \nonumber \\
  & \qquad \qquad = \frac{\partial}{\partial x_j} \int d\vec{y}\,\,y_i \, C(\vec x, \vec y, \vec x, t) \nonumber \\
  & \qquad \qquad = \frac{\partial}{\partial x_j} \int d\vec{y}\,\, y_i \, \text{Prob}(\vec x \to \vec y, t) \nonumber \\
  & \qquad \qquad = \frac{\partial}{\partial x_j} \langle x_i(t) \rangle\,,
  \label{}
\end{align}
where $\text{Prob}(\vec x \to \vec y, t) = |\langle \vec y| U |\vec x\rangle|^2$ is the probability density of transitioning from $\vec{x}$ to $\vec{y}$ in time $t$ and $\langle x_i(t) \rangle$ is average $i$th coordinate at time $t$ with the initial condition $x_i(0) = x_i$. 

The propagator representation of OTO operators provides a complementary perspective to the Wigner phase space representation.  While the propagator representation does not have a manifest classical limit, it does show how the OTO operator can be expressed by objects (namely, derivatives of the first moment of the chaos kernel) with intuitive classical analogs. 

\section*{Spin Degrees of Freedom}
Next we repeat our above analyses for spin degrees of freedom.  For motivation, we review a classical spin system. In particular, we consider a top with angular momentum $\vec J$ driven by a Hamiltonian expressible in terms of the components of $\vec J$. Then, $\abs{\vec J}^2$ is conserved so that our dynamics are effectively on the sphere $S^2$ with coordinates $(\theta, \varphi)$. For $\abs{\vec J} = \sqrt{s(s+1)}$, the Poisson bracket is given by
\begin{equation}
  \left\{ f,g \right\}_{\text{Poisson}}  = \frac{1}{\sqrt{s(s+1)}\sin \theta} \left( \frac{\partial f}{\partial \varphi} \frac{\partial g}{\partial \theta} - \frac{\partial f}{\partial \theta} \frac{\partial g}{\partial \varphi} \right)\,.
  \label{}
\end{equation}
Now we consider the sensitivity of the $z$-component of the angular momentum with respect to the initial azimuthal angle:
\begin{equation}
  \frac{\partial J^z(t)}{\partial \varphi_0} = \frac{\partial }{\partial \varphi_0}\,\left(\sqrt{s(s+1)}\cos \theta(t)\right).
  \label{}
\end{equation}
In terms of Poisson brackets this is
\begin{equation}
  \left\{ J^z(t), J^z(0) \right\}_{\text{Poisson}} .
  \label{}
\end{equation}
Hence, a spin OTO operator has the form
\begin{equation}
\label{spinOTOop1}
- [\widehat J^z(t) , \widehat J^z(0)]^2. 
\end{equation}
While our analysis extends straightforwardly to other types of commutators of spin operators, we restrict our attention on the OTO operator in \ceqref{spinOTOop1}.

\subsection*{Spins via Phase Space Representation}
The Wigner phase space formulation can be extended to spins~\cite{varilly1989moyal,klimov2002moyal,koczor2016time}. Although our final result for the spin OTO operator is straightforward, we will need to make many definitions to apply the spin phase space formalism.  Let $\widehat A$ be an operator which acts on the Hilbert space $\C^{2s+1}$ of a spin-$s$  particle.  The spin Weyl correspondence associates to each operator $\widehat A$ a phase space function $\mathcal{W}_{\widehat A} : S^2 \to \C$ defined by
\begin{equation}
  \mathcal{W}_{\widehat A}(\theta,\varphi) \equiv \tr\left( \widehat A \, \widehat w_{\mathcal{W}} (\theta,\varphi)\right)
  \label{}
\end{equation}
where
\begin{equation}
  \widehat w_{\mathcal{W}}(\theta,\varphi) \equiv \frac{2 \sqrt{\pi}}{\sqrt{2s+1}} \sum_{\ell =0}^{2s}\sum_{m=-\ell }^{\ell} Y_{\ell m}^*(\theta,\varphi) \widehat T_{\ell m}^{(s)},
  \label{}
\end{equation}
In the equation above, $Y_{\ell m}$ are the spherical harmonics and $\widehat T_{\ell m}^{(s)}$ are the irreducible tensor operators~\cite{varshalovich1988quantum}
\begin{equation}
  \widehat T_{\ell m}^{(s)} \equiv \sqrt{\frac{2\ell +1}{2s+1}} \sum_{m,m'=-s}^{s} C_{sm, \ell m}^{sm'} \ket{s,m'}\bra{s,m}\, ,
  \label{}
\end{equation}
where $C_{sm, \ell m}^{sm'}$ are Clebsch-Gordan coefficients:
\begin{equation}
  C_{j_1 m_1, j_2 m_2}^{J m} \equiv \left(\bra{j_1 \, m_1} \ot \bra{j_2\, m_2}\right) \ket{J m}.
  \label{}
\end{equation}
It will be useful to decompose $\widehat{A}$ into irreducible tensor operators as
\begin{equation}
  \widehat A = \sum_{\ell =0}^{2s} \sum_{m=-\ell }^{\ell } A_{\ell m} \widehat T_{\ell m}^{(s)}\,.
  \label{}
\end{equation}
We denote by $\deg(\widehat A)$ the largest $\ell $ such that $A_{\ell m} \neq 0$ for some $m$. This is called the \textit{degree of non-linearity} of $\widehat{A}$.

The spin Weyl correspondence induces the spin analog of the Moyal star product. It is given by
\begin{align}
  &\mathcal{W}_{\widehat A} \star \mathcal{W}_{\widehat B} \equiv  \sqrt{2s+1} \sum_{j=0}^{j_{\max}} \frac{(-1)^j}{j!(2s+j+1)!} \nonumber\\
  & \quad \times \widetilde F^{-1} (\mathcal{L}^2) \left[\left(S^{+(j)} \widetilde F (\mathcal{L}^2) W_{\widehat A} \right)\left(S^{-(j)} \widetilde F (\mathcal{L}^2) W_{\widehat B} \right)\right]
  \label{spinStar}
\end{align}
where $j_{\max} \equiv \min\{\deg(\widehat A), \deg(\widehat{B})\}$, $\mathcal{L}^2$ is the Casimir operator on the sphere
\begin{align}
  & \mathcal{L}^2 \equiv -\left[ \frac{\partial^2}{\partial \theta^2} + \cot \theta \frac{\partial}{\partial \theta} + \frac{1}{\sin^2 \theta} \frac{\partial^2}{\partial \varphi^2} \right] \\
  & \mathcal{L}^2 Y_{\ell,m} = \ell(\ell+1) Y_{\ell,m},
  \label{}
\end{align}
$\widetilde F$ is a function such that 
\begin{equation}
  \widetilde F(\mathcal{L}^2) Y_{\ell ,m} \equiv \sqrt{(2s+\ell +1)!(2s-\ell )!} \, Y_{\ell ,m}\,,
  \label{}
\end{equation}
and the $S^\pm$ are differential operators defined by
\begin{equation}
  S^{\pm (j)} \equiv 
  \begin{cases}
    \prod_{m=0}^{j-1} \left( m \cot \theta - \frac{\partial}{\partial \theta} \mp \frac{i}{\sin \theta} \frac{\partial}{\partial \varphi} \right) & j>0 \\
    1 & j=0
  \end{cases}\,.
  \label{}
\end{equation}

We can now use the formalism to obtain a semiclassical expansion.  Applying the spin Weyl correspondence to the spin OTO operator, we obtain
\begin{align}
\left\{\!\!\left\{W_{\widehat J^z}(t) , W_{\widehat J^z}(0) \right\}\!\!\right\}^{\star 2} = \left( \frac{\partial \mathcal{W}_{\hat J^z}(t)}{\partial \varphi} \right)^{\star 2},
 \label{}
\end{align}
where the bracket $\left\{\!\left\{ \cdot,\cdot \right\}\!\right\}$ is defined as in the spatial case, and the equality follows from generalizing Theorem 5 in~\cite{varilly1989moyal}. 
We can now use \ceqref{spinStar} to obtain
\begin{widetext}
  \begin{equation}
    \left\{\!\!\left\{W_{\widehat J^z}(t) , W_{\widehat J^z}(0) \right\}\!\!\right\}^{\star 2} = \sqrt{2s+1} \sum_{j=0}^{\deg(\widehat J^z(t))} \frac{(-1)^j}{j!(2s+j+1)!} \, \widetilde F^{-1}(\mathcal{L}^2) \left(\left( S^{+(j)} \widetilde F (\mathcal{L}^2) \frac{ \partial W_{\widehat J^z}(t)}{\partial \varphi} \right)\left( S^{-(j)} \widetilde F (\mathcal{L}^2) \frac{ \partial W_{\widehat J^z}(t)}{\partial \varphi} \right)\right).
    \label{spinOTOresult1}
  \end{equation}
\end{widetext}
Note that similar to the spatial case, $W_{\widehat J^z}(t)$ can be expanded in $1/s$, where the leading term is the solution to the classical equations of motion.  However, again the corrections depend on the specific dynamics, and so we absorb them into $W_{\widehat J^z}(t)$.  Also note that the number of terms in \ceqref{spinOTOresult1} is directly related to non-linearity of the operator $\widehat J^z(t)$. This implies higher derivative corrections are induced by chaotic dynamics that increase $\deg(\widehat J^z (t))$, which is initially unity at $t=0$. (Non-linearity can be phrased in terms of an expansion in a basis of operators. For a discussion about the time dependence of such expansions, see Appendix \ref{OpGrowth}.)  This is analogous to the spatial case where chaotic dynamics make $X_i(t)$ nonlinear in $\vec{x}$ and $\vec{p}$, and thereby generate higher derivative corrections in the OTO operator in \ceqref{OTOexpansion1}.

\ceqref{spinOTOresult1} has a particularly enlightening form in the semiclassical limit where $s \gg 1$. In this limit, we can approximate the star product using the small expansion parameter $\varepsilon \equiv \frac{1}{2s+1}$~\cite{klimov2002moyal}:
\begin{align}
  & \mathcal{W}_{\widehat A} \star  \mathcal{W}_{\widehat B}\nonumber\\
  & =  \int_0^{2\pi} \frac{d \psi}{2 \pi} \, W_{\widehat A} \,\exp\left[ \frac{\varepsilon}{2} \left( \cev S_- \vec S_+ - \cev S_+ \vec S_- \right) \right] W_{\widehat B} + \mathcal{O}(\varepsilon^3),
  \label{spinStarApprox}
\end{align}
where $(\theta,\varphi,\psi)$ are the Euler angles, and 
\begin{equation}
  S_{\pm} \equiv i e^{\mp i\psi} \left( \pm \cot \theta \frac{\partial}{\partial \psi} + i \frac{\partial}{\partial\theta} \mp \frac{1}{\sin \theta} \frac{\partial}{\partial \varphi} \right).
  \label{}
\end{equation}
In the limit of large $s$, we can use \ceqref{spinStarApprox} to expand \ceqref{spinOTOresult1} to second order in $\varepsilon$:
\begin{align}
  & \left\{\!\!\left\{  W_{\widehat J^z}(t) , W_{\widehat J^z}(0)\right\}\!\!\right\}^{\star 2} \nonumber\\
  & \quad =  \left( \frac{\partial W_{\widehat J^z}(t)}{\partial \varphi} \right)^2 - \varepsilon^2 \det \mathrm{Hess}_{S^2}\left(  \frac{\partial W_{\widehat J^z}(t)}{\partial \varphi}\right) + \mathcal{O}(\varepsilon^3)\,,
  \label{spinExpansion1}
\end{align}
\vskip1in
\noindent where $\mathrm{Hess}_{S^2} \equiv \nabla \nabla$ is the Hessian tensor on the sphere. In the $(\wh \theta, \wh \varphi)$ basis, $\nabla = \frac{\partial}{\partial\theta} \wh \theta + \frac{1}{\sin \theta} \frac{\partial}{\partial \varphi} \wh \varphi\,$. Hence we intriguingly find that \ceqref{spinExpansion1} has precisely the same form as \ceqref{OTOexpansion2}, even up to constant factors.  Furthermore, using the above scrambling time analysis and replacing $\hbar$ with $1/(2s+1)$, we expect the spin OTO operator to saturate at times $t_{\text{scr}} \sim \log(2s+1)$ for large $s$.
\vskip.4in
\subsection*{Spins via Propagator Representation}

We now consider a more general system with $N$ spin-$s$ particles and use propagators to express the spin OTO operators $- [\widehat J^z_i(t), \widehat J^z_j ]$, where $\widehat J^z_i$ is the $z$-component of the spin of the $i$th particle.  We will find that the spin OTO operator can be interpreted as the sensitivity of a quantity similar to $\mu_i$ defined in \ceqref{1stMom}.

To define a propagator for spins, we use spin coherent states which are defined as~\cite{varilly1989moyal}
\begin{equation}
  \ket{s, \textbf{n}} \equiv \sum_{m=-s}^{s} \binom{2s}{s+m}^{\frac{1}{2}} \cos^{s+m} \frac{\theta}{2} \sin^{s-m} \frac{\theta}{2} e^{-im \varphi} \ket{s,m},
  \label{}
\end{equation}
where $s$ is the spin quantum number, $\textbf{n} \in S^2$ and $\theta,\varphi$ are its polar and azimuthal angles, respectively. The $\ket{s,m}$ are eigenvectors of $J^z$ with eigenvalues $m$. Note that the spin coherent states form a complete basis, thereby allowing us to insert resolutions of the identity:
\begin{equation}
  I = \frac{2s+1}{4\pi} \int_{S^2} d \textbf{n}\,\state{s, \textbf{n}} \,.
  \label{}
\end{equation}
We denote the spin propagator by
\begin{equation}
  K^s(\vec{\textbf{n}}, \vec{\textbf{m}}, t) \equiv \bra{s, \vec{\textbf{m}}} U \ket{s, \vec{\textbf{n}}},
  \label{}
\end{equation}
where $\ket{s, \vec{\textbf{n}}} \equiv \bigotimes_{a=1}^N \ket{s, \textbf{n}_i}$.  Then, we define the \textit{spin chaos kernel} for the $i$th particle as
\begin{align}
 & C^s_i(\vec{\textbf{n}}, \vec{\textbf{m}}, \overline{\varphi}_{\textbf{m}_i}, \vec{\textbf{k}},t) \nonumber \\ &\quad = \int_{0}^{2\pi}  d\varphi_{\textbf{m}_i}K^s(\vec{\textbf{n}}, \vec{\textbf{m}},t)\,\frac{\partial}{\partial \varphi_{\textbf{m}_i}}(K^s)^*(\vec{\textbf{k}}, \vec{\textbf{m}}, t),
  \label{spinChaosKernel}
\end{align}
where $\overline{\varphi}_{\textbf{m}_i}$ means $C^s_i$ is \textit{not} a function of $\varphi_{\textbf{m}_i}$. Geometrically, $C^s_i$ is the complex area enclosed by the loop traced out by $(K^s(\vec{\textbf{n}}, \vec{\textbf{m}}, t), (K^s)^*(\vec{\textbf{k}}, \vec{\textbf{m}},t)) \in \C^2$ as $\varphi_{\textbf{m}_i}$ varies from $0$ to $2\pi$.  \ceqref{spinChaosKernel} is analogous to the chaos kernel defined in \ceqref{cKernel} and we can likewise define a quantity analogous to the first moment of $\mu_i$ defined in  \ceqref{1stMom} by integrating $C_i^s$ over all the degrees of freedom of $\vec{\textbf{m}}$ other than $\varphi_{\textbf{m}_i}$:
\begin{align}
  & \mu_i^s(\vec{\textbf{n}}, \vec{\textbf{k}}, t) \nonumber \\
  &= \left( \frac{2s+1}{4\pi} \right)^N  \int \prod_{\ell\neq i} d\textbf{m}_\ell \, \sin \theta_{\textbf{m}_i} d \theta_{\textbf{m}_i} C^s_i(\vec{\textbf{n}}, \vec{\textbf{m}}, \overline{\varphi}_{\textbf{m}_i}, \vec{\textbf{k}},t)\nonumber \\
  & = \left( \frac{2s+1}{4\pi} \right)^N \int d\vec{\textbf{m}}\, K^s(\vec{\textbf{n}}, \vec{\textbf{m}},t) \frac{\partial}{\partial \varphi_{\textbf{m}_i}} (K^s)^*(\vec{\textbf{k}}, \vec{\textbf{m}},t).
  \label{}
\end{align}
$\mu_i^s(\vec{\textbf{n}}, \vec{\textbf{k}}, t)$ can be interpreted as the average sensitivity of the amplitude to go back in time to the angles $\vec{\textbf{k}}$ given that the initial angles were $\vec{\textbf{n}}$.

Putting everything together, the spin OTO operator is given by
\begin{widetext} 
\begin{align}
  & -[\wh J_i^z(t) ,\wh  J_j^z]^2 = - \left(\frac{2s+1}{4\pi}\right)^{4N} \int \left(\prod_{\ell=1}^4 d\textbf{n}_\ell\right) \ket{\vec{\textbf{n}}_1}\bra{\vec{\textbf{n}}_4}  \left. \frac{\partial}{\partial \varphi_{21,j}^+} \mu^s_i(\vec{\textbf{n}}_2, \vec{\textbf{n}}_1,t) \right \vert_{\varphi_{21,j}^-} f^s(\vec{\textbf{n}}_2, \vec{\textbf{n}}_3)\left. \frac{\partial}{\partial \varphi_{43,j}^+} \mu^s_i(\vec{\textbf{n}}_4,\vec{\textbf{n}}_3,t) \right\vert_{\varphi_{43,j}^-},
  \label{spinOTOkernel}
\end{align}
\end{widetext}
where $\varphi_{ab,j}^\pm \equiv \frac{1}{2} ( \varphi_{\left( \textbf{n}_a \right)_j} \pm \varphi_{\left( \textbf{n}_b \right)_j})$ and $f^s(\vec{\textbf{n}}, \vec{\textbf{m}}) \equiv \langle s, \vec{\textbf{n}} \vert s, \vec{\textbf{m}} \rangle$. For $s=\frac{1}{2}$, for instance, we have
\\
\begin{align}
  & f^{\frac{1}{2}}(\vec{\textbf{n}},\vec{\textbf{m}}) = \prod_{l=1}^N \bigg( \sin \frac{\theta_{\textbf{n}_l}}{2}\sin \frac{\theta_{\textbf{m}_l}}{2} e^{-\frac{i}{2} \left( \varphi_{\textbf{n}_l} - \varphi_{\textbf{m}_l} \right)} \nonumber \\
  & \qquad \qquad \qquad \quad + \cos\frac{\theta_{\textbf{n}_l}}{2}\cos\frac{\theta_{\textbf{m}_l}}{2} e^{\frac{i}{2} \left( \varphi_{\textbf{n}_l} - \varphi_{\textbf{m}_l} \right)}\bigg).
  \label{FullSpinOTOprop}
\end{align}
We observe
\begin{equation}
  (\mu^s_i)^*(\vec{\textbf{n}}_2, \vec{\textbf{n}}_1,t) = -\mu^s_i(\vec{\textbf{n}}_1, \vec{\textbf{n}}_2,t)
  \label{}
\end{equation}
by integrating by parts and noticing that the boundary term vanishes by periodicity. Furthermore,
\begin{equation}
  (f^s)^*(\vec{\textbf{n}},\vec{\textbf{m}}) = f^s(\vec{\textbf{m}},\vec{\textbf{n}}),
  \label{}
\end{equation}
so \ceqref{FullSpinOTOprop} is manifestly Hermitian.

We see that \ceqref{spinOTOkernel} has a similar form as \ceqref{OTOkernel} and admits a similar interpretation.  We find again that the OTO operator has essentially the same structure across spatial and spin degrees of freedom.

\section*{Discussion}

We have shown how the OTO operator reflects sensitivity to initial conditions beyond the classical limit.  In particular, for both spatial and spin degrees of freedom, the quantum corrections to the OTO operator form a tower of higher derivatives with respect to the initial conditions of a time-evolved observable.  The Wigner phase space formalism is central to this analysis, and provides a clear way of detailing semiclassical expansions in either $\hbar$ or $1/(2s + 1)$.  Our analysis has provided an estimate and geometric interpretation of the scrambling time, which agrees with known results and has a clearer interpretation in some respects. Additionally, we have recast OTO operators in terms of propagators, and have found that they reflect the sensitivity of the propagation of the system along a spacetime arc when its initial and final points are varied. All of these analyses can be extended to quantum field theories~\cite{curtright1999wigner}.

Perhaps what is most striking is that the OTO operators for both spatial and spin degrees of freedom have such similar form in their phase space representations.  For example, the leading terms in both expansions have the same form, as do the leading higher derivative corrections which are Hessians. This universality of form is not obvious in other treatments of the OTO operator in which semiclassical expansions are not manifest.  The unifying ingredient is phase space, which is a natural setting for both classical \textit{and} quantum chaos~\cite{korsch1981evolution,hutchinson1980quantum,greenbaum2005semiclassical,bracken2006semiquantum,maia2008semiclassical,chaudhury2009quantum,haake2013quantum}.  This suggests it may be interesting to explore other recent developments in quantum chaos using the phase space formalism.
\\ \\
\noindent \textbf{Acknowledgements.\quad}JC is supported by the Fannie
and John Hertz Foundation and the Stanford Graduate Fellowship program.
DD is supported by the Stanford Graduate Fellowship program.  We would like to thank Xingshan Cui, Jonathan Dowling, Nicole Younger-Halpern, Patrick Hayden, Edward Mazenc, Stephen Shenker, Joseph V\'{a}rilly, Cosmas Zachos and Wojciech Zurek for valuable discussions and feedback. 

\bibliographystyle{apsrev4-1}
\bibliography{Ref}

\cleardoublepage
\onecolumngrid
\appendix

\section{OTO Operator to Second Order}
\label{sec:manyWignerOTO}
Recall that $Y_{ij}(t) \equiv \frac{\partial X_i(t)}{\partial x_j}$.  To second order, we can write Eqn. \!(\ref{OTOexpansion1}) as
\begin{align}
 &\{\!\{ X_i(\vec{x},\vec{p},t), P(\vec{x},\vec{p},0) \}\!\}^{\star 2} \nonumber \\
 &\approx \left( Y_{ij}(t) \right)^2 - \frac{\hbar^2}{4}\left( \sum_{k=1}^{d\,N} \left(\frac{\partial^2 Y_{ij}(t)}{\partial p_k^2} \frac{\partial^2 Y_{ij}(t)}{\partial x_k^2} - \left(\frac{\partial^2 Y_{ij}(t)}{\partial x_k \partial p_k}\right)^2\right) + 2 \sum_{\substack{k, k'=1 \\ k \not = k'}}^{d\,N} \left( \frac{\partial^2 Y_{ij}(t)}{\partial p_k \partial p_{k'}}\frac{\partial^2 Y_{ij}(t)}{\partial x_k \partial x_{k'}} - \frac{\partial^2 Y_{ij}(t)}{\partial x_k \partial p_{k'}}\frac{\partial^2 Y_{ij}(t)}{\partial p_k \partial x_{k'}} \right)\right) \nonumber \\
  & = \left( Y_{ij}(t) \right)^2 - \frac{\hbar^2}{4} \left( \sum_{k,k'=1}^{d\,N} \underset{2k,2k'}{{\det}_2} \text{Hess}(Y_{ij}(t))\right)
  \label{manyWignerOTO}
\end{align}
where $\underset{i,j}{\det_2}$ denotes the principal minor of order 2 with $(i,j)$ being the indices of the bottom right entry.  We can visualize the $\mathcal{O}(\hbar^2)$ term by partitioning the $2\,d\,N\times 2\,d\,N$ Hessian matrix into $2\times 2$ submatrices in the natural way: \\
\begin{equation}
  \text{Hess}\left( Y_{ij}(t) \right) = \left[
  \begin{array}{c|c|c}
    H_{1,1} & \cdots & H_{1, d\,N}\\
    \hline
    H_{2,1} & \cdots & H_{2, d\,N}\\
    \hline
    \vdots & \ddots & \vdots\\
    \hline
    H_{d\,N,1} & \cdots & H_{d\,N, d\,N}
  \end{array}
  \right].
  \label{}
\end{equation}
\\
\noindent Then, the leading higher derivative correction is proportional to the sum of the determinants of all the $2\times 2$ submatrices. Note that \ceqref{manyWignerOTO} manifestly simplifies to \ceqref{OTOexpansion2} in the case of a single particle in one dimension.

\section{Comments on Operator Growth}
\label{OpGrowth}
Consider a lattice spin system of $N$ spin-$1/2$ particles (each on their own site) evolving with some unitary time evolution.  In this system, two operators localized to different sites will commute.  For example, the angular momentum operators $\wh J_i^z$ and $\wh J_j^z$ which act on sites $i$ and $j$ (for $i \not = j$) commute: $[\wh J_i^z, \wh J_j^z] = 0$.  However, working in the Heisenberg picture, we see that in contrast $[\wh J_i^z(t), \wh J_j^z(0)]$ does \textit{not} necessarily equal zero, and in fact will equal a sum of operators that will spread across all sites as a function of time until it saturates at some distribution over all operators.
The reason is because $\wh J_i^z(t)$ has an expansion of the form
\begin{align}
\label{HeisenbergEvol1}
\wh J_i^z(t) &= a(t) \,\mathds{1} + \sum_{i=1}^N \sum_{\alpha \in \left\{ 0,x,y,z \right\}} b_{i}^{\alpha}(t) \, \wh J_i^\alpha + \sum_{i,j=1}^N \sum_{\alpha, \beta \in \left\{ 0,x,y,z \right\}} c_{ij}^{\alpha \beta}(t) \, \wh J_i^\alpha \wh J_j^\beta + \sum_{i,j,k=1}^N \sum_{\alpha, \beta, \gamma \in \left\{ 0,x,y,z \right\}} d_{ijk}^{\alpha \beta \gamma}(t) \, \wh J_i^\alpha \wh J_j^\beta \wh J_k^\gamma + \cdots
\end{align}
where $\wh J_i^0 \equiv \wh{\mathds{1}}$, and so $\wh J_i^z(t)$ spreads across the space of operators as it evolves in time, eventually overlapping more and more with $\wh J_j^z(0) = \wh J_j^z$.  So $[\wh J_i^z(t), \wh J_j^z(0)]$ and related commutators provide a natural measure of quantum chaos, since their time dependence captures how local degrees of freedom are spreading in the system.

Often, we want to compute quantities like $\text{tr}(\rho\, [\wh J_i^z(t), \wh J_j^z(0)])$ for some state $\rho$, such as a Gibbs state.  However, for many states $\rho$ we expect that there will be lots of cancellation in $\text{tr}(\rho\, [\wh J_i^z(t), \wh J_j^z(0)])$, since in light of Eqn.\! (\ref{HeisenbergEvol1}), all of the terms not proportional to the identity can have expectation values which are either positive \textit{or} negative, and their time-dependent coefficients can also be positive or negative.  As remarked in the introduction, the cancellation obscures the operator growth under time evolution, and to remedy this, we instead consider $\text{tr}(\rho\, [\wh J_i^z(t), \wh J_j^z(0)]^2)$ which is manifestly positive.

In the spin example, the salient feature which characterizes quantum chaos is the growth of operators under time evolution. However in the case of the $-\frac{1}{\hbar^2}[\wh{x}(t), \wh{p}(0)]^2$ operator, writing
\begin{equation}
\wh{x}(t) = \sum_{m,n} c_{mn}(t) \, (\wh{x}^m \wh{p}^n + \wh{p}^n \wh{x}^m)
\end{equation}
where the $c_{mn}(t)$ are real, we find that
\begin{equation}
\label{ApproxOTO1}
-\frac{1}{\hbar^2}\text{tr}(\rho\,[\widehat{x}(t),\widehat{p}]^2) \approx\sum_{m,n} m^2\,c_{mn}(t)^2  \,\text{tr}(\rho\,(\widehat{x}^{m-1}\widehat{p}^n + \widehat{p}^n \widehat{x}^{m-1})^2)
\end{equation}
where ``$\approx$" signifies that we have dropped terms which are not manifestly positive that may cancel out.  From \ceqref{ApproxOTO1}, we see that the OTO operator captures how much the $\widehat{x}(t)$ operator is spreading. There is an $m^2$ weighting each term that contains a $\widehat{x}^{m-1}$; this increases the rate of growth of the expectation of the OTO operator as $\widehat{x}(t)$ spreads to operators containing $\widehat{x}^m$\,'s for progressively larger values of $m$.


\end{document}